\documentclass[lettersize,journal]{IEEEtran}
\usepackage{amsmath,amsfonts}
\usepackage{algorithm}
\usepackage{array}
\usepackage[caption=false,font=normalsize,labelfont=sf,textfont=sf]{subfig}
\usepackage{textcomp}
\usepackage{stfloats}
\usepackage{url}
\usepackage{verbatim}
\usepackage{graphicx}
\usepackage{cite}
\usepackage[most]{tcolorbox}
\usepackage{xcolor}         
\usepackage{colortbl}
\usepackage{tabularx}
\usepackage{makecell}
\usepackage{multirow}
\usepackage{algpseudocode}
\newcommand{\best}{\cellcolor{blue!20}}  
\usepackage[switch,columnwise]{lineno} 
\newcommand{\revision}[1]{\textcolor{black}{#1}}

\definecolor{mybg}{RGB}{217, 234, 211}   
\definecolor{myborder}{RGB}{106, 168, 79}  

\newcounter{finding}

\newtcolorbox{findingbox}[1][]{%
  enhanced,
  colback=mybg!70,      
  colframe=myborder!80, 
  coltitle=black,
  fonttitle=\bfseries,
  boxrule=0.8pt,
  arc=0mm,
  left=1mm,
  right=1mm,
  top=1mm,
  bottom=1mm,
  sharp corners=south,
}

\usepackage{booktabs}       
\usepackage{amsfonts}       
\usepackage{microtype}      
\usepackage{amsmath}
\usepackage{graphicx}
\usepackage{enumitem}
\usepackage{amssymb}

\usepackage{colortbl}
\definecolor{iccvblue}{rgb}{0.21,0.49,0.74}

\usepackage[pagebackref,breaklinks,colorlinks,allcolors=iccvblue]{hyperref}
\begin{document}

\title{How Far Are We from Generating Missing Modalities with Foundation Models?}

\author{Guanzhou Ke, Bo Wang, Guoqing Chao, Weiming Hu,~\IEEEmembership{Senior Member,~IEEE}, Shengfeng He,~\IEEEmembership{Senior Member,~IEEE}
\thanks{This research is supported by the Guangdong Natural Science Funds for Distinguished Young Scholars (Grant 2023B1515020097), the National Research Foundation Singapore under the AI Singapore Programme (AISG Award No: AISG4-TC-2025-018-SGKR), and the Lee Kong Chian Fellowships. Corresponding author: Shengfeng He.}
\thanks{Guanzhou Ke is with the Institute of Data Science and Intelligent Decision Support, Beijing Jiaotong University, Beijing 100080, China. He is also with the School of Computing and Information Systems, Singapore Management University, Singapore 178903, Singapore. E-mail: guanzhouk@gmail.com}%
\thanks{Bo Wang and Weiming Hu are with the State Key Laboratory of Multimodal Artificial Intelligence Systems, Institute of Automation, Chinese Academy of Sciences, Beijing 100190, China. E-mail: wangbo@ia.ac.cn, wmhu@nlpr.ia.ac.cn}
\thanks{Guoqing Chao is with the School of Computer Science and Technology, Harbin Institute of Technology, Weihai 264209, China. E-mail: guoqingchao@hit.edu.cn}
\thanks{Shengfeng He is with the School of Computing and Information Systems, Singapore Management University, Singapore 178903, Singapore. E-mail: shengfenghe@smu.edu.sg}

}

\markboth{IEEE Transactions on Pattern Analysis and Machine Intelligence}%
{Guanzhou Ke \MakeLowercase{\textit{et al.}}: How Far Are We from Generating Missing Modalities with Foundation Models?}


\maketitle

\begin{abstract}
Multimodal foundation models have demonstrated impressive capabilities across diverse tasks. However, their potential as plug-and-play solutions for missing modality reconstruction remains underexplored. 
To bridge this gap, we identify and formalize three potential paradigms for missing modality reconstruction, and perform a comprehensive evaluation across these paradigms, covering 42 model variants in terms of reconstruction accuracy and adaptability to downstream tasks.
Our analysis reveals that current foundation models often fall short in two critical aspects: (i) \textit{fine-grained semantic extraction from the available modalities}, and (ii) \textit{robust validation of generated modalities}. These limitations lead to suboptimal and, at times, misaligned generations.
To address these challenges, we propose an agentic framework tailored for missing modality reconstruction. This framework dynamically formulates modality-aware mining strategies based on the input context, facilitating the extraction of richer and more discriminative semantic features. In addition, we introduce a \textit{self-refinement mechanism}, which iteratively verifies and enhances the quality of generated modalities through internal feedback. Experimental results show that our method reduces FID for missing image reconstruction by at least 14\% and MER for missing text reconstruction by at least 10\% compared to baselines. Code is released at: \url{https://github.com/Guanzhou-Ke/AFM2}.

\end{abstract}

\section{Introduction}

\IEEEPARstart{M}{issing} modalities are common in real-world multimodal systems, often due to sensor failures or privacy constraints. This challenge is particularly pronounced in domains such as healthcare and robotics, where certain modalities may be difficult or impossible to acquire. Most existing methods are limited in two key aspects: (i) they primarily focus on recovering high-level semantic representations of missing data, rather than reconstruct the raw modality itself~\cite{wang2023incomplete,ma2021smil,ma2022multimodal,zhang2022m3care}. While effective for certain tasks, this limits applications that require low-level modality inputs, such as image-conditioned generation or speech synthesis; (ii) they typically require retraining or fine-tuning on domain-specific data, resulting in poor generalization and limited transferability across diverse scenarios. A promising alternative is to leverage large multimodal models (LMMs)~\cite{wang2024emu3, chen2025janus, hurst2024gpt, xu2025qwen2, xie2024show} to reconstruct missing modalities, owing to their strong generalization capabilities. Unfortunately, unlike conventional approaches, fine-tuning LMMs for each downstream reconstruction task can be prohibitively expensive in terms of both computation and data. As a compromise, designing a training-free paradigm for missing modality reconstruction becomes an attractive direction~\cite{ke2025knowledge}. However, this approach remains largely underexplored, which leads us to ask: \textit{How far are we from reconstructing missing modalities using foundation models in a training-free manner?}


\begin{figure}[t]
  \centering
  \includegraphics[width=1\linewidth]{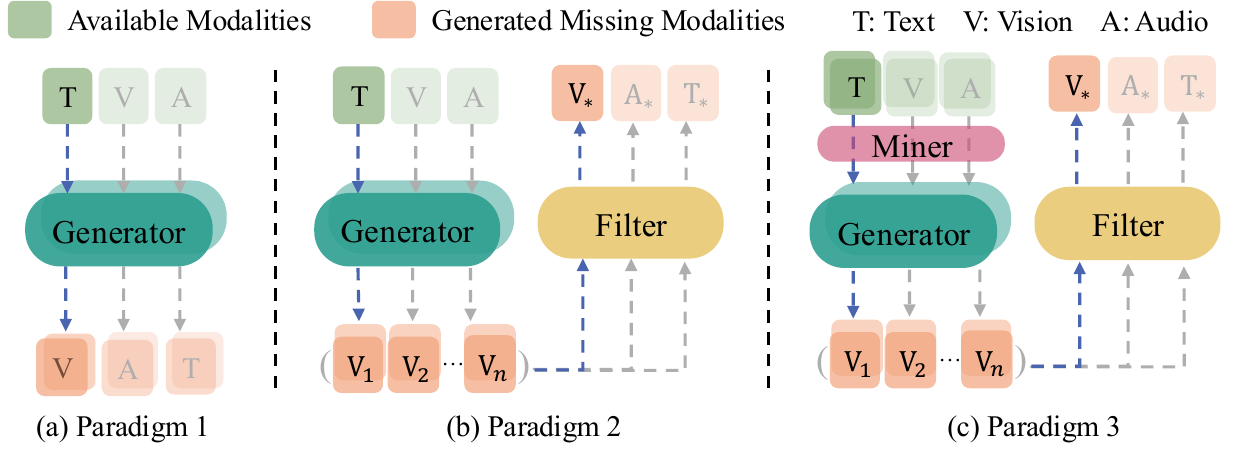}
  \caption{
  Overview of three paradigms for missing modality generation. 
  (a) \textit{Direct Generation}: the model directly generates the missing modality from the available inputs. 
  (b) \textit{Generation with Filter}: candidate outputs are first generated, and the most semantically aligned one is selected by a filter. 
  (c) \textit{Generation with Miner and Filter}: a cross-modal miner first mines and integrates multimodal knowledge before generation, which is followed by filtering to select the optimal output.
  }
  \label{fig:paradigm}
\end{figure}

To systematically investigate the challenge of missing modality generation using LMMs, we decompose the reconstruction process into three essential components: a \textit{generator}, a \textit{miner}, and a \textit{filter module}. The generator synthesizes the missing modality; the miner extracts informative cross-modal cues from the observed inputs; and the filter module selects the output that best aligns with the intended semantics among multiple candidates. As illustrated in Fig.~\ref{fig:paradigm}, this functional decomposition enables us to categorize existing methods into three paradigms. \revision{This design is connected to recent agentic and evaluator-based frameworks. The miner role is related to task planning and tool coordination in HuggingGPT~\cite{shen2023hugginggpt}, but is specialized here for extracting fine-grained cross-modal evidence from observed modalities. The filter module, instantiated as a verifier in AFM$^2$, is related to MLLM-as-a-Judge~\cite{chen2024mllm}, but is further embedded into a self-refinement loop to guide re-generation when candidates are semantically inconsistent.} Overall, this formulation is motivated by the structure of the problem rather than by ad hoc design choices: in a training-free setting, successful reconstruction requires not only generation, but also explicit evidence mining and semantic verification at test time.

The first paradigm adopts direct conditional generation (e.g., text-to-image), where the generator (such as Stable Diffusion~\cite{rombach2022high}) is directly prompted with the available modality~\cite{reed2016generative, ramesh2022hierarchical, yu2022scaling, liu2023audioldm}. However, these methods lack mechanisms for reliable semantic evaluation or verification, often resulting in outputs that deviate from the intended semantics. Therefore, the second paradigm extends the first by generating multiple outputs and leveraging a filter model (e.g., ImageBind~\cite{girdhar2023imagebind}) to select the most coherent one. While this improves output fidelity to some extent, it typically ignores fine-grained cues embedded in observed modalities, such as object counts and visual attributes.
The third paradigm introduces a modality mining step, where a foundation model (e.g., GPT-4o~\cite{hurst2024gpt}) infers latent semantic relationships across modalities to guide generation~\cite{guo2025can, gal2024comfygen, cong2025can, ke2025knowledge}. While these paradigms are thus constructed by progressively integrating the core components we have just outlined, it remains unclear how each component individually contributes to the reconstruction quality, or to what extent LMMs can perform training-free missing modality generation. Understanding these questions is critical to revealing the upper bounds of LMMs capabilities in this setting.


To this end, we construct 42 model variants spanning three paradigms and conduct extensive experiments, which reveal two critical limitations in current foundation model-based approaches. First, audio generation performance lags significantly behind that of image and text, primarily due to an over-reliance on text-based prompts that lack multimodal grounding. Incorporating a dedicated miner module to extract richer semantic context substantially narrows this gap. Second, most existing methods lack a robust mechanism for verifying generated outputs, often resulting in semantic inconsistencies. Introducing a filter module to assess and verify generated candidates notably improves alignment and cross-modal coherence.

Motivated by these findings, we propose an \textbf{A}gentic \textbf{F}ramework for \textbf{M}issing \textbf{M}odality generation (\textbf{AFM$^2$}), which integrates three collaborative agents: a \textit{miner}, a \textit{verifier}, and a \textit{generator}. The miner adaptively extracts fine-grained elements such as objects, actions, and spatiotemporal cues, tailored to the input modality. The verifier synthesizes guidance by filtering noisy or inconsistent content and estimating the semantic plausibility of candidate generations. The generator then selects appropriate models and candidate counts based on this refined guidance. To further enhance robustness, we introduce a \textit{self-refinement mechanism}: if no candidate meets the verifier’s quality threshold, the guidance is refined and the generation process is restarted in an iterative loop. Comprehensive evaluation across all 42 variants demonstrates that our framework, particularly the miner and verifier agents, yields substantial improvements in the quality, consistency, and reliability of missing modality generation over competitive baselines. Our main contributions are summarized as follows:

\begin{itemize}[leftmargin=0.5cm]
    \item We construct and evaluate 42 model variants across three paradigms, revealing audio as the most challenging modality and highlighting the importance of mining and filter modules.
    \item We identify key limitations in current approaches, including weak audio grounding due to text-only prompting and the absence of output verification, and show that targeted architectural interventions can significantly improve generation quality.
    \item We introduce an agentic framework with self-refinement that dynamically coordinates mining, verification, and generation, improving semantic consistency and robustness across modalities.
\end{itemize}

\section{Related Work}

\subsection{Missing Modality Generation} 

Missing modality generation aims to reconstruct absent modalities based on available ones. Early approaches rely on statistical imputation (e.g., zero~\cite{parthasarathy2020training} or mean filling~\cite{zhang2020deep}), while later methods~\cite{wang2023multi, ma2021smil, lian2023gcnet, pham2019found} focus on recovering high-level semantic representations to avoid the complexity of raw data reconstruction.
More recent work leverages large pre-trained models for richer, more expressive outputs~\cite{lee2023multimodal, guo-etal-2024-multimodal, ke2025knowledge, meng2024multi, xiao2024fgc2f}. For example, Ke et al.~\cite{ke2025knowledge} employ pre-trained diffusion models guided by structured knowledge graphs to synthesize missing modalities, mainly in vision-language settings. In contrast, our agentic framework introduces a self-refinement mechanism that ensures output quality without relying on handcrafted knowledge, making it broadly compatible with diverse multimodal backbones.


\subsection{Generative Foundation Models} 

Generative foundation models are capable of producing high-quality content, such as text~\cite{yang2024qwen2, hurst2024gpt, touvron2023llama}, images~\cite{rombach2022high, ramesh2022hierarchical, flux2024}, or audio~\cite{evans2025stable, liu2023audioldm, liu2024audioldm}, from either noise or prompts in a creative manner. These models are typically categorized into three main types: autoregressive models~\cite{wang2024emu3, hurst2024gpt, wu2024next}, diffusion models~\cite{ramesh2022hierarchical, evans2025stable, liu2024audioldm, rombach2022high}, and generative adversarial networks (GANs)~\cite{goodfellow2020generative, mirza2014conditional, brock2018large}. In particular, conditioning the generation process~\cite{mirza2014conditional, ho2022classifier, ramesh2022hierarchical} on specific inputs has been shown to improve output quality and better align results with user intent. In recent years, a growing body of work has explored the use of such pre-trained generative models for missing modality generation~\cite{meng2024multi, kebaili2025amm, ouyang2023missdiff}, achieving promising results in domains such as healthcare~\cite{meng2024multi, xiao2024fgc2f} and recommendation systems~\cite{li2025generating}. However, existing approaches often focus on unimodal missing scenarios and lack mechanisms for mining relevant information from available modalities or ranking generated candidates. In contrast, our method introduces two key agents: an automated miner agent that extracts cross-modal information, and a robust verifier agent that evaluates and refines generation outputs. Together, they significantly enhance the quality of missing modality generation.

\subsection{Agentic Artificial Intelligence}

Agentic AI~\cite{durante2024agent} leverages sophisticated reasoning and iterative planning to autonomously solve complex, multi-step problems. In recent years, it has achieved remarkable success in domains such as computer operator~\cite{agashe2024agent, sager2025ai}, mathematics~\cite{zhu2022solving, liang2023encouraging, wang2023math}, and programming~\cite{roychoudhury2025agentic, white2024building}. In this work, we focus on training-free agentic AI, where foundation models act as planners, executors, and evaluators without task-specific fine-tuning.

Pioneering works like Toolformer~\cite{schick2023toolformer} and HuggingGPT~\cite{shen2023hugginggpt} demonstrate how LLMs can solve complex problems by invoking external tools. Among them, HuggingGPT presents a structured agentic paradigm encompassing task planning, model selection, and tool coordination, which is closely related to our miner agent that extracts fine-grained cross-modal evidence from observed modalities. \revision{Recent MLLM-as-a-Judge studies further show that multimodal models can serve as evaluators to score, compare, or rank candidate outputs according to semantic alignment and factual groundedness~\cite{chen2024mllm}, which is closely related to our verifier agent.}

This line of work has inspired follow-up studies such as NavGPT~\cite{zhou2024navgpt}, which introduces explicit reasoning into vision-and-language navigation, and AdaptAgent~\cite{verma2024adaptagent}, which leverages multimodal few-shot demonstrations for rapid adaptation in web-based tasks. Inspired by these advances, we formulate missing modality generation as a multi-step agentic process. We employ large multi-modal models to reason over observed modalities, mine cross-modal relationships, invoke appropriate generators, and verify outputs in an iterative loop. To the best of our knowledge, this is the first work to apply an Agentic AI paradigm to missing modality generation.

\section{Evaluating the Potential Paradigms}
\label{sec:evaluating}

The task of missing modality generation addresses the common and challenging problem of reconstructing a missing modality from a set of available, observed ones. For instance, consider a scenario where a textual description and an audio clip are available, but the corresponding image is inaccessible due to factors such as transmission errors or privacy constraints. Such data incompleteness can severely hinder downstream tasks that rely on a full set of modalities. A promising remedy, therefore, is to generate the missing image while ensuring semantic consistency with the provided text and audio. While traditional approaches often require training specialized models and may generalize poorly across scenarios, the recent success of large-scale foundation models has inspired a new research direction toward training-free missing modality generation. However, a fundamental question remains underexplored: how capable are these powerful models when used directly, without any task-specific fine-tuning? This work explores this question by systematically constructing and evaluating common paradigms built upon foundation models.

\subsection{Overview}
We propose that a foundation model-based framework for missing modality prediction can be decomposed into three key components: a \textit{generator}, a \textit{miner}, and a \textit{filter module}. The generator produces the missing modality given an instruction; the miner extracts contextual cues from observed modalities to guide generation; and the filter module selects the most semantically aligned output from multiple candidates. Our goal is to systematically investigate different paradigms and assess the individual contribution of each component.

This decomposition is intended as a functional abstraction of training-free missing-modality generation rather than an ad hoc categorization. Effective reconstruction in this setting requires not only synthesizing the target modality, but also exposing fine-grained evidence from observed modalities and verifying whether generated candidates are semantically faithful to the multimodal context. \revision{From the perspective of prior agentic systems, the miner resembles a task-specific planning and evidence-extraction module, analogous to the planning component in HuggingGPT~\cite{shen2023hugginggpt}, while the filter/verifier instantiates a multimodal judging mechanism related to MLLM-as-a-Judge~\cite{chen2024mllm}. Separating these roles allows us to analyze their individual contributions and compare different foundation model-based paradigms systematically.}

Following a bottom-up design, we construct three paradigms by incrementally integrating these components (Fig.~\ref{fig:paradigm}):
(a) Paradigm 1 uses only the generator for direct generation;
(b) Paradigm 2 adds a filter module to select the best candidate;
(c) Paradigm 3 further incorporates a miner to mine cross-modal relationships and enhance prompt informativeness.

To systematically evaluate the effectiveness of different paradigms and individual components under a training-free setting, we select two state-of-the-art foundation models for each module to ensure diversity, as shown in Table~\ref{tab:components}. Specifically, we adopt six pretrained generative models to reconstruct the missing image, text, and audio. For example, we leverage Stable Diffusion 3.5 (SD3.5) and FLUX.1 dev to generate missing images based on the input modality (text). For the filter module, we consider two strategies: (i) ImageBind~\cite{girdhar2023imagebind} which measures the similarity between generated and observed modalities, and (ii) MLLM-as-a-Judge~\cite{chen2024mllm} which employs Qwen2.5-Omni-7B~\cite{xu2025qwen2} to assess the semantic quality of generated outputs\footnote{The prompt template used for MJ is provided in the appendix.}. To extract contextual cross-modal cues, we adopt two pretrained models as miners: Qwen2.5-Omni-7B (QO)~\cite{xu2025qwen2} and OpenAI GPT-4o (4o). Based on the structure of each paradigm, we systematically combine these modules as shown in Table~\ref{tab:paradigm_overview}. To ensure fair comparisons, we control all variables except the one under investigation. This results in a total of 42 baseline variants. In the following sections, we evaluate each variant from two perspectives: (i) the quality of the generated missing modalities, and (ii) the performance impact on downstream tasks. This allows us to quantify the role and effectiveness of each individual component.



\begin{table}[t]
  \centering
  \caption{Components used for modality generator, miner, and filter.}
  \renewcommand{\arraystretch}{1.2}
  \setlength{\tabcolsep}{2pt}
  \begin{tabular}{lll}
    \toprule
    \multicolumn{3}{c}{\textbf{\textit{Generator}}} \\
    \midrule
    \textbf{Image} & \textbf{Text} & \textbf{Audio} \\
    Stable Diffusion 3.5~\cite{rombach2022high} & Qwen2.5-VL-7B~\cite{bai2025qwen2} & AudioLDM 2~\cite{liu2024audioldm} \\
    FLUX.1 dev~\cite{flux2024} & Qwen2.5-Omni-7B~\cite{xu2025qwen2} & Stable Audio 1.0~\cite{evans2025stable} \\
    \midrule
    \multicolumn{3}{c}{\textbf{\textit{Miner}}} \\
    \midrule
    \textbf{Model} & \textbf{Flag} & \textbf{Modalities} \\
    Qwen2.5-Omni-7B~\cite{xu2025qwen2} & QO & Vision, Language, Audio \\
    OpenAI GPT-4o~\cite{hurst2024gpt} & 4O & Vision, Language \\
    \midrule
    \multicolumn{3}{c}{\textbf{\textit{Filter}}} \\
    \midrule
    \textbf{Strategy} & \textbf{Flag} & \textbf{Backbone} \\
    ImageBind~\cite{girdhar2023imagebind} & IB & ImageBind \\
    MLLM-as-a-Judge~\cite{chen2024mllm} & MJ & Qwen2.5-Omni-7B \\
    \bottomrule
  \end{tabular}
  \label{tab:components}
\end{table}

\begin{table}[t]
  \centering
  \caption{Overview of all paradigm settings.}
  \renewcommand{\arraystretch}{1.2}
  \setlength{\tabcolsep}{7pt}
  \begin{tabularx}{\linewidth}{lp{1.5cm}ccc}
    \toprule
    Paradigm & Generator & Miner & Filter & Variants \\
    \midrule
    Paradigm 1 & \multirow{7}[1]{*}{\makecell[l]{All generators\\declared in \\ Table \ref{tab:components}}} & / & / & 6 \\
    Paradigm 2 (IB) &  & / & IB & 6 \\
    Paradigm 2 (MJ) &  & / & MJ & 6 \\
    Paradigm 3 (QO) &  & QO & MJ & 6 \\
    Paradigm 3 (4o) &  & 4o & MJ & 6 \\
    Paradigm 3 (QO+IB) &  & QO & IB & 6 \\
    Paradigm 3 (4o+IB) &  & 4o & IB & 6 \\
    \bottomrule
  \end{tabularx}
  \label{tab:paradigm_overview}
\end{table}

\subsection{Experiment Setting}
\label{sec:exp-set}


\subsubsection{Datasets Construction}
We conduct experiments on four datasets: VGGSound, MSRVTT, AudioCaps, and COCO2014. The first three are preprocessed into tri-modal samples containing image, text, and audio, while COCO2014 is used as a bi-modal image-text dataset. This setup enables evaluation under both tri-modal and bi-modal missing-modality scenarios. Detailed dataset statistics and preprocessing procedures are provided in the appendix. Unless otherwise specified, all experiments follow a single-missing setting, where one modality is masked as the target and the remaining modalities are used as observed inputs for reconstruction.

\subsubsection{Evaluation Metrics}
We employ metrics that assess both the fidelity and semantic consistency of the generated modalities. For evaluating image generation, we use Fréchet Inception Distance (FID)~\cite{heusel2017gans} and CLIP image similarity (CLIP-I)~\cite{radford2021learning}.  
For evaluating text generation, we report Match Error Rate (MER)~\cite{morris2004and} and CLIP text similarity (CLIP-T)~\cite{radford2021learning}.  
For evaluating audio generation, we adopt Scale-Invariant Signal-to-Noise Ratio (SI-SNR)~\cite{luo2018tasnet} and Perceptual Evaluation of Speech Quality (PESQ)~\cite{recommendation2001perceptual}. CLIP-I and CLIP-T measure the semantic similarity between generated and ground-truth modalities using CLIP-based embeddings. To evaluate downstream task performance, we report classification metrics including F1 score and average precision (AP) on the corresponding test sets.  
Additional details on metric computation are provided in the appendix.

\subsubsection{Implementation Settings} We evaluate foundation models for missing modality prediction from two perspectives: (i) generation accuracy, and (ii) adaptability to downstream tasks such as classification. All experiments are conducted on eight NVIDIA L40 GPUs, each with 46GB of memory.

\textbf{Protocol of missing modality generation.}
Unless otherwise specified, all experiments follow a single-missing modality setting: for each sample, only one target modality is masked as missing, and all remaining modalities are treated as observed inputs for reconstructing the missing one. 
For example, on tri-modal datasets such as VGGSound, MSRVTT, and AudioCaps, when text is missing, the model uses image and audio as inputs to generate the missing text; similarly, when image (or audio) is missing, the remaining two modalities are used as conditions. 
For COCO2014, which contains image and text modalities, the model reconstructs one modality from the other. For generation-quality evaluation, the missing rate controls the proportion of test samples selected for masking, whereas for downstream-task evaluation it controls the proportion of training samples requiring completion.



\textbf{Generation Accuracy.}
To assess generation quality, we simulate missing-modality scenarios on the test set with missing rates of 30\%, 50\%, and 70\%. Here, the missing rate denotes the proportion of test samples in which one target modality is masked for reconstruction, rather than the degree of information removal within each individual sample. This setting is used to simulate different levels of missing-modality prevalence in deployment scenarios. Since the evaluated paradigms are training-free, varying the missing rate does not change the reconstruction mechanism itself; instead, it changes the composition and scale of the masked test subset on which generation quality is evaluated. For each configuration, a model reconstructs the missing modality based on the available inputs. The predicted outputs are then compared against the ground-truth targets, and we report FID and CLIP-I for image, MER and CLIP-T for text, and PESQ and SI-SNR for audio.

\textbf{Downstream Classification.}
To assess downstream utility, we simulate missing data on the training set. The missing modalities are predicted using each model, and the resulting completed training data is used to train a classifier. We then evaluate classification performance on the original test set. This protocol measures how well foundation models can enhance downstream tasks within their existing knowledge structures. Unless otherwise specified, we adopt ImageBind’s encoder as the frozen modality encoder and attach a trainable multi-layer perceptron (MLP) for classification. Detailed training hyperparameters are provided in the appendix.

\subsection{Quantitative Analysis}

\begin{figure*}[htbp]
  \centering
   \includegraphics[width=1\linewidth]{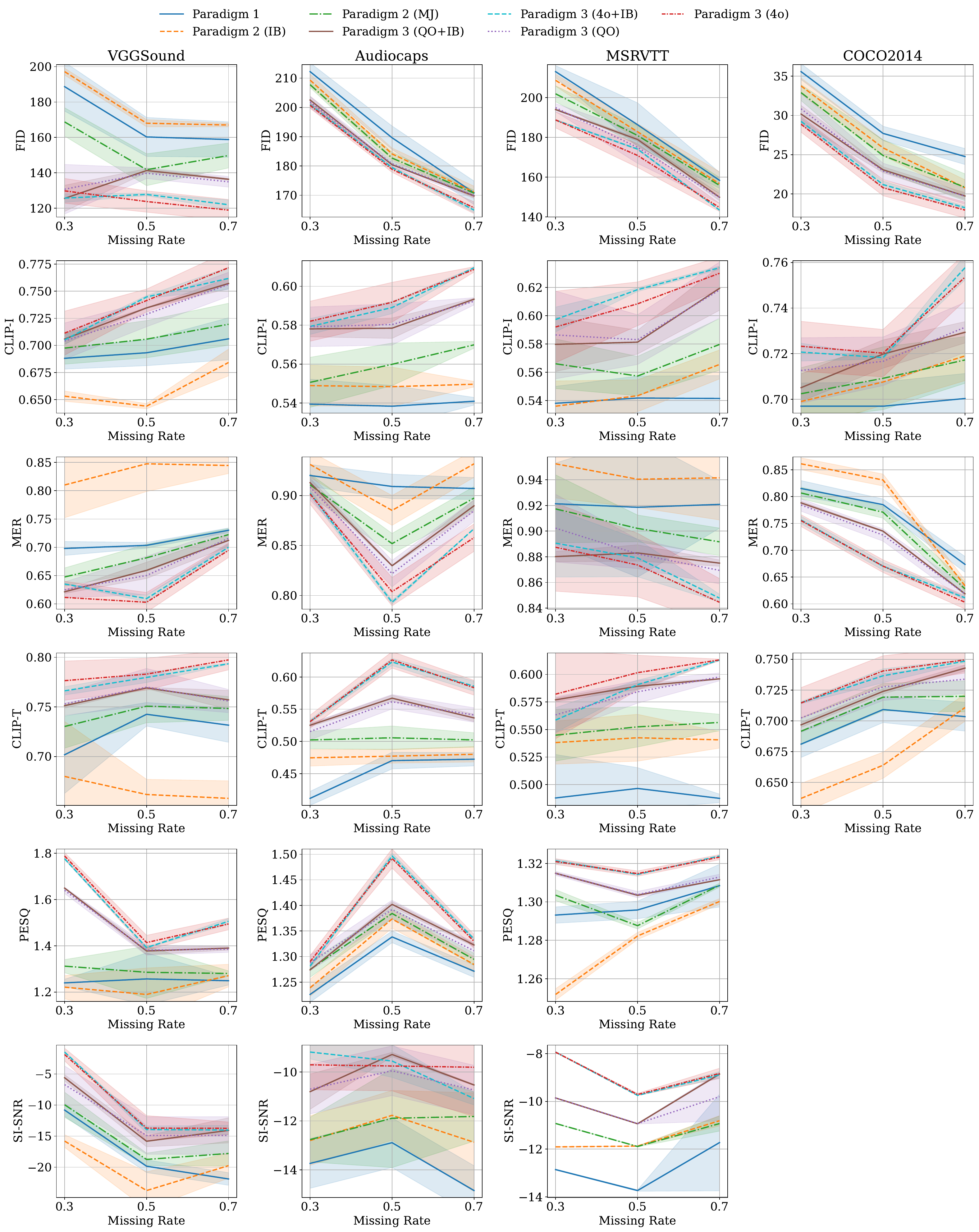}
   \caption{The major quantitative results of the three paradigms across four datasets. For missing vision generation, we FID ($\downarrow$) and CLIP-I ($\uparrow$) similarity. For missing text generation, we use MER ($\downarrow$) and CLIP-T ($\uparrow$) similarity. For missing audio generation, we report PESQ ($\uparrow$) and SI-SNR ($\uparrow$).}
   \label{fig:comprehensive}
\end{figure*}

\begin{figure*}[t]
  \centering
   \includegraphics[width=1\linewidth]{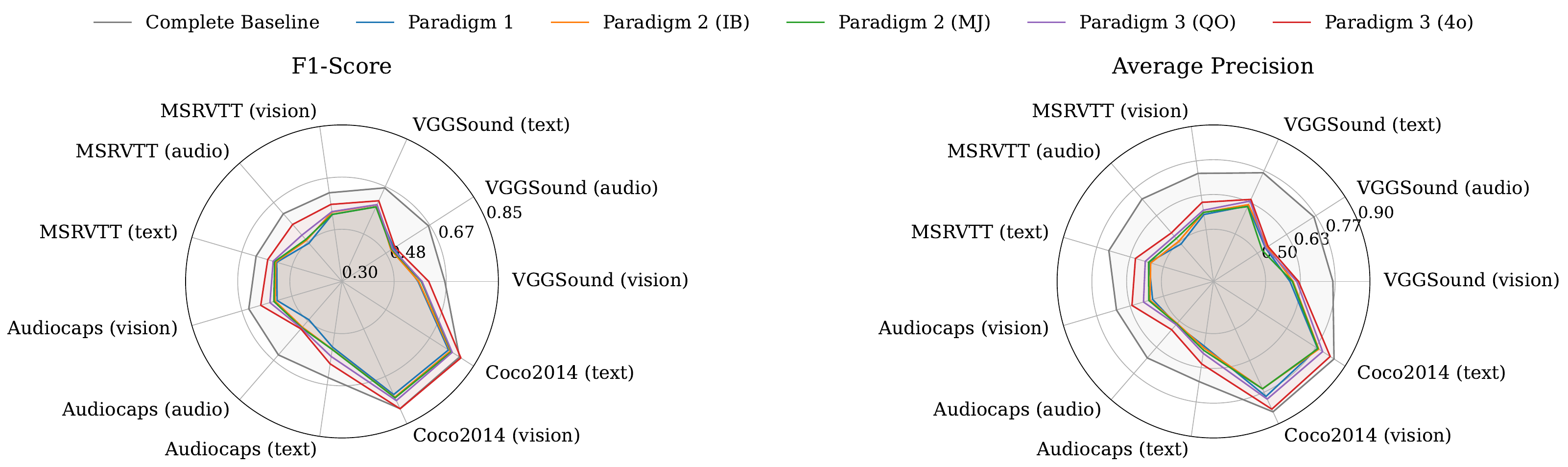}
   \caption{Comparison of F1-score and average precision (AP) across four datasets for all paradigms under a 70\% missing modality rate.}
   \label{fig:radio}
\end{figure*}

We report the major quantitative results of the three paradigms across four datasets, as summarized in Figure~\ref{fig:comprehensive} and Figure~\ref{fig:radio}.

\subsubsection{For Missing Modality Generation} 
To facilitate comparison, we group the 42 baseline configurations into seven representative setting categories and report their aggregate performance. Here, ``aggregate performance'' denotes the mean±std computed over the variants belonging to each setting category. This aggregation is adopted to provide a more focused comparison of the overall strengths and weaknesses of different setting categories, while the complete fine-grained results of all individual variants are provided in the appendix. The first is Paradigm 1, which directly uses the generator to predict missing modalities. The second and third are Paradigm 2 (IB) and Paradigm 2 (MJ), which incorporate ImageBind (IB) and MLLM-as-a-Judge (MJ) as filter modules, respectively. The remaining baselines are Paradigm 3 (QO), Paradigm 3 (4o), Paradigm 3 (QO+IB), and Paradigm 3 (4o+IB) that further integrate Qwen2.5-Omni-7B and GPT-4o as modality miners on top of Paradigm 2. We report the generation results using the evaluation metrics defined in Section~\ref{sec:exp-set}.

As shown in Fig.~\ref{fig:comprehensive}, both Paradigm 2 and Paradigm 3 consistently outperform Paradigm 1 across all four datasets, confirming the effectiveness of incorporating filtering and mining modules in improving generation quality. Within Paradigms 2 and 3, the MJ-based variants (\textcolor{green}{green} line) outperform their IB-based counterparts (\textcolor{orange}{orange} line) in several cases—particularly for image and text generation on VGGSound and AudioCaps—demonstrating the superior generalization ability of more flexible filtering strategies. On the other hand, among all Paradigm 3 variants, Paradigm 3 (4o) (\textcolor{red}{red} line), which employs GPT-4o as the miner, achieves the highest overall performance. This observation suggests that stronger cross-modal reasoning capability can more effectively enhance the quality of missing modality generation, a trend that is also consistent with previous findings~\cite{ke2025knowledge}.

\revision{A notable exception is observed for text-generation MER on COCO and image-generation FID across datasets, where higher missing rates sometimes lead to better scores. This should be interpreted cautiously because, in our protocol, the missing rate controls the proportion of test samples selected for masking, rather than the amount of information removed from each individual sample. Thus, different missing rates correspond to different evaluated subsets, while the reconstruction mechanism remains unchanged. For COCO text generation, the trend may be related to the in-domain nature of COCO for current LMMs and the lexical sensitivity of MER; results on the original COCO captions in Appendix Table~IV show a similar tendency, suggesting that caption rewriting is not the sole cause. For image generation, FID is a distribution-level metric and can vary with subset composition, diversity, and sample size across datasets. We therefore regard the apparent improvement at higher missing rates as a subset-composition and metric-sensitivity effect, rather than evidence that increased missingness intrinsically improves generation quality.}

However, all variants exhibit negative SI-SNR scores for audio generation, indicating a substantial gap between generated and real audio. We attribute this to two primary factors:
(i) the performance gap between audio generators and their counterparts for other modalities, and
(ii) the inherent complexity of audio synthesis, particularly in our datasets, which include diverse environmental sounds (e.g., car horns, ambient noise). Such sounds are difficult to infer from other modalities due to their weak semantic alignment.
This highlights missing audio generation as a particularly challenging problem in the multimodal generation landscape.

\begin{table*}[t]
  \centering
  \caption{Impact of the number of generation candidates on the VGGSound dataset. Cells highlighted in \colorbox{blue!20}{blue} indicate the best performance. Time represents the average latency per sample in seconds. For GPT-4o, communication overhead is ignored, and local deployment latency is used as a reference.}

  \renewcommand{\arraystretch}{1.2}
  \setlength{\tabcolsep}{2pt}
  {\begin{tabular}{cc|cccccc|cccccc}
  \toprule
  & Missing Rate & \multicolumn{6}{c|}{0.3} & \multicolumn{6}{c}{0.7}  \\
       & Missing Type & \multicolumn{3}{c}{Image} & \multicolumn{3}{c|}{Text} & \multicolumn{3}{c}{Image} & \multicolumn{3}{c}{Text} \\
  \cmidrule{3-14}
  N of Gene. & Metrics & FID $\downarrow$ & CLIP-I $\uparrow$ & Time (s) & MER $\downarrow$ & CLIP-T $\uparrow$ & Time (s) & FID $\downarrow$ & CLIP-I $\uparrow$ & Time (s) & MER $\downarrow$ & CLIP-T $\uparrow$ & Time (s) \\
  \midrule
  \multirow{4}[0]{*}{N = 5} 
      & Paradigm 2 (IB) & 197.17 & 0.65 & \best 17.47 & 0.81 & 0.67 & 5.54 & 167.01 & 0.68 & \best 17.71 & 0.84 & 0.65 &  5.53 \\
      & Paradigm 2 (MJ) & 168.75 & 0.69 & 21.30 & 0.64 & 0.73 & \best 5.46 & 149.76 & 0.71 & 20.63 & 0.72 & 0.74 & \best 5.47 \\
      & Paradigm 3 (QO) & 130.71 & 0.70 & 23.59 & 0.62 & 0.75 & 7.96 & 134.91 & 0.75 & 22.77 & 0.71 & 0.75 & 7.90 \\
      & Paradigm 3 (4o) & \best 129.79 & \best 0.71 & 23.80 & \best 0.61 & \best 0.77 & 8.16 & \best 118.91 & \best 0.77 & 23.37 & \best 0.69 & \best 0.79 & 8.05 \\
  \midrule
  \multirow{4}[0]{*}{N=15} 
      & Paradigm 2 (IB) & 113.57 & 0.68 & \best 52.59 & 0.76 & 0.69 & \best 16.43 & 97.16 & 0.71 & \best 53.90 & 0.71 & 0.68 & \best 16.70 \\
      & Paradigm 2 (MJ) & 101.39 & 0.71 & 63.09 & 0.61 & 0.73 & 17.28 & 80.27 & 0.78 & 61.34 & 0.63 & 0.75 & 17.50 \\
      & Paradigm 3 (QO) & 83.15 & 0.78 & 65.34 & 0.53 & 0.76 & 19.42 & 72.94 & \best 0.79 & 63.76 & 0.44 & \best 0.81 & 19.62 \\
      & Paradigm 3 (4o) & \best 72.11 & \best 0.79 & 65.88 & \best 0.51 & \best 0.81 & 19.76 & \best 61.26 & \best 0.79 & 64.32 & \best 0.42 & \best 0.81 & 19.98 \\
  \midrule
  \multirow{4}[0]{*}{N=25} 
      & Paradigm 2 (IB) & 102.06 & 0.69 & \best 87.84 & 0.61 & 0.71 & \best 22.41 & 97.02 & 0.74 & \best 87.13 & 0.56 & 0.73 & \best 22.48 \\
      & Paradigm 2 (MJ) & 81.34 & 0.71 & 105.22 & 0.65 & 0.72 & 24.70 & 79.22 & 0.78 & 103.68 & 0.56 & 0.77 & 24.36 \\
      & Paradigm 3 (QO) & 75.13 & 0.75 & 107.21 & 0.56 & 0.78 & 29.68 & 72.11 & 0.78 & 106.02 & 0.43 & 0.81 & 29.85 \\
      & Paradigm 3 (4o) & \best 63.27 & \best 0.79 & 107.80 & \best 0.47 & \best 0.81 & 30.13 & \best 62.19 & \best 0.79 & 106.52 & \best 0.39 & \best 0.82 & 30.28 \\
  \bottomrule
  \end{tabular}}%
  \label{tab:test-time-scaling}%
\end{table*}

\subsubsection{For Downstream Task Evaluation} To further evaluate the effectiveness of different paradigms on downstream tasks, we report classification performance on four datasets under a 70\% missing rate, using F1-score and average precision (AP) as metrics (see Fig.~\ref{fig:radio}). 


The performance trends in Fig.~\ref{fig:radio} closely mirror those observed in Fig.~\ref{fig:comprehensive}: both Paradigm 2 and Paradigm 3 outperform Paradigm 1, and the GPT-4o-based variant of Paradigm 3 achieves the best overall performance across all datasets. Notably, on the COCO2014 dataset, Paradigm 3 (4o) achieves classification accuracy that nearly matches the performance of a baseline trained with the complete data. This suggests that Paradigm 3 (4o) provides strong semantic reconstructions that are comparable to original data—a phenomenon similar to data augmentation, as also discussed in prior work~\cite{liu2024best}.

\begin{findingbox}
\label{find-1}
\textbf{Takeaway:} Incorporating candidates filtering strategies and cross-modal mining with strong models greatly improves missing modality generation quality and enables downstream performance comparable to using complete data.
\end{findingbox}

\subsection{The Impact of Generation Quantity on Generation Quality}

We further investigate whether increasing the number of generation candidates at inference time can enhance the quality of missing modality generation. For missing images, we sample $N$ outputs using the same generation prompt, while for missing text, we generate $N$ alternative candidates. Table~\ref{tab:test-time-scaling} presents results on the VGGSound dataset for three candidate counts ($N = 5, 15, 25$), evaluated by FID, CLIP-I, MER, CLIP-T, and average inference time. Additional results are provided in the appendix.

The experimental results in Table~\ref{tab:test-time-scaling} show that model performance consistently improves as the number of candidates increases. For example, at a missing rate of 0.3, increasing the number of candidates for Paradigm 2 (IB) from $N=5$ to $N=25$ reduces the FID from 197.17 to 102.06, an improvement of 48.23\%. The average inference time also increases by approximately four times (from 17.47 seconds to 87.84 seconds), but this is still less than the fivefold increase in candidate count. Notably, Paradigm 3 (4o) outperforms all other baselines in nearly every setting. For instance, in the image modality with a missing rate of 0.3, its FID drops from 129.79 at $N=5$ to 72.11 at $N=15$, and further to 63.27 at $N=25$. In the text modality, the MER decreases from 0.61 to 0.47, while the CLIP-T score rises from 0.77 to 0.81. These findings suggest that increasing the number of generation candidates leads to higher quality generated modalities, which is consistent with observations in previous work~\cite{yue2025does}. Overall, increasing the number of generation candidates can significantly improve missing modality generation quality in training-free settings, but it also introduces considerable computational overhead. Therefore, how to balance the trade-off between the number of candidates and computational cost remains an open problem.


\begin{findingbox}
\label{find-2}
\textbf{Takeaway:} Increasing the number of generation candidates at inference time substantially improves the quality of missing modality generation in training-free settings.
\end{findingbox}

\begin{table*}[htbp]
  \centering
  \caption{Impact of information mining granularity on the VGGSound dataset under a 70\% missing rate. Cells highlighted in \colorbox{blue!20}{blue} indicate the best performance for each metric.}
    \renewcommand{\arraystretch}{1.2}
    {\begin{tabular}{cc|c|c|c|c|c|c|c}
    \toprule
    \multicolumn{9}{c}{\textit{For Missing Image}} \\
    \toprule
    Granularity & \multicolumn{2}{c|}{Baseline} & \multicolumn{2}{c|}{Object} & \multicolumn{2}{c|}{Object [\textit{location}]} & \multicolumn{2}{c}{Object[\textit{color}]} \\
   \cmidrule{2-9}       & FID $\downarrow$   & CLIP-I $\uparrow$& FID $\downarrow$  & CLIP-I $\uparrow$& FID $\downarrow$  & CLIP-I $\uparrow$ & FID $\downarrow$  & CLIP-I $\uparrow$ \\
   \midrule
     Paradigm 3 (QO) & 134.91 & 0.75  & 103.33 & 0.78  & 94.48 & 0.81  & 101.46 & 0.79 \\
    Paradigm 3 (4o) & 118.91 & 0.77  & 97.19 & 0.81  & \best 88.62 & \best 0.83  & 98.52 & 0.81 \\
    \toprule
    \multicolumn{9}{c}{\textit{For Missing Text}} \\
    \toprule
          & MER $\downarrow$  & CLIP-T $\uparrow$ & MER $\downarrow$  & CLIP-T $\uparrow$ & MER $\downarrow$  & 
          CLIP-T $\uparrow$ & MER $\downarrow$  & CLIP-T $\uparrow$ \\
    \midrule
    Paradigm 3 (QO) & 0.71  & 0.75  & 0.63  & 0.77  & 0.64  & 0.78  & 0.63  & 0.76 \\
    Paradigm 3 (4o) & 0.69  & 0.79  & 0.57  & 0.81  & \best 0.53  & \best 0.83  & 0.55  & 0.81 \\
    \bottomrule
    \end{tabular}}%
  \label{tab:granularity}%
\end{table*}%

\subsection{Impact of Information Mining Granularity}


To assess how mining granularity influences missing modality generation, we conduct experiments on the VGGSound dataset with a 70\% missing rate. Starting from a baseline that uses default prompts and $N=5$ candidates, we gradually enhance the miner module to extract: (i) objects only, (ii) objects with spatial locations, and (iii) objects with color attributes. Full prompt templates are provided in the appendix.

As shown in Table~\ref{tab:granularity}, we find that explicitly specifying the type of information to be mined can substantially improve the quality of missing modality generation. For example, comparing the baseline with the variant that mines object information, the FID decreases from 118.91 to 97.19, representing an 18\% improvement. This demonstrates that clarifying the specific information to be extracted enables the miner module to perform more effectively, which in turn improves generation quality. Furthermore, adding finer-grained guidance such as location or color information leads to even greater gains. For instance, including location details reduces the FID from 118.91 to 88.62 and the MER from 0.69 to 0.53. While incorporating color provides only marginal improvements over location alone, it still outperforms the baseline, highlighting the benefit of detailed semantic guidance.



\begin{findingbox}
\label{find-3}
\textbf{Takeaway:} The granularity and quality of information mined from the observed modalities are positively correlated with the quality of missing modality generation.
\end{findingbox}

These results suggest that the proposed generator--filter--miner decomposition is not merely conceptual, but reflects the functional requirements of training-free missing-modality generation. While the generator provides the basic synthesis capability, the filter improves semantic reliability through candidate selection, and the miner further enhances reconstruction by exposing fine-grained cross-modal cues that are difficult to exploit through direct prompting alone.
\section{Agentic Missing Generation Framework}
\label{sec:afm}

\subsection{Overview}

\begin{figure*}[t]
  \centering
   \includegraphics[width=1\linewidth]{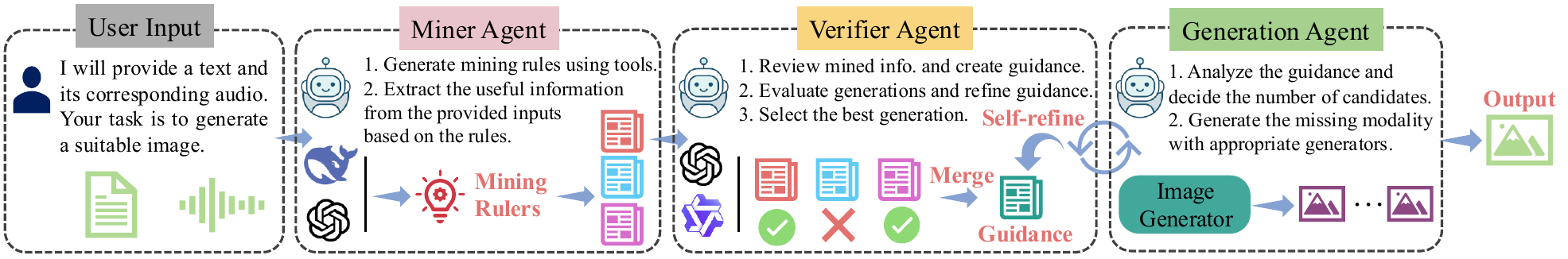}
   \vspace{-5mm}\caption{Overview of an agentic framework for generating missing modalities.}\vspace{-2mm}
   \label{fig:framework}
\end{figure*}

Building on the preceding empirical findings, we argue that an effective framework for missing modality generation should not only support automatic fine-grained information mining from the input and robust generation verification and filtering, but also generate a larger number of candidates more efficiently. To this end, we propose an \textbf{A}gentic \textbf{F}ramework for \textbf{M}issing \textbf{M}odality (AFM$^2$), whose workflow is illustrated in Fig.~\ref{fig:framework}. AFM$^2$ consists of three collaborative agents: a \textit{miner agent}, a \textit{verifier agent}, and a \textit{generation agent}. The miner agent extracts key contextual cues from the available modalities. The verifier agent evaluates and refines the generation guidance through a self-refinement mechanism and ranks the generated candidates to select the most plausible output. The generation agent dynamically invokes appropriate generators and determines the number of candidates to produce based on the refined guidance. Detailed implementations of each agent are described in the following sections.

\subsection{Agent Design}

\subsubsection{Miner Agent} The miner agent automatically generates mining rules and extracts semantic cues from available modalities. It uses a reasoning model (e.g., DeepSeek-R1~\cite{guo2025deepseek} or OpenAI’s o3) to infer task-specific rules that identify key elements such as objects, scenes, actions, and spatiotemporal relations. These rules are adaptively tailored to the input modality (e.g., text, image, or audio) and used to guide generation. For instance, given text (“a dog is playing in a park”) and audio (“a dog’s barking”), it may extract the scene “a dog barking in a park.” We use DeepSeek-R1 for rule inference and Qwen2.5-Omni-7B for information extraction by default.

\subsubsection{Verifier Agent} The verifier agent evaluates the information extracted by the miner agent and is responsible for generating and refining the generation guidance. It first assesses the quality of the mined content, filters out irrelevant or redundant information, and summarizes a compact guidance text that describes potential objects, environments, actions, and other semantic cues. After the generation agent produces multiple candidates, the verifier scores each output based on cross-modal consistency, semantic similarity, and hallucination risk. By default, each candidate is rated on a scale from 1 to 5. If all scores fall below a predefined threshold, the verifier triggers a self-refinement process: it revises the generation guidance based on the evaluation feedback and initiates a new generation cycle. This creates a quality control feedback loop that mitigates hallucinations and semantic misalignments. By default, we use Qwen2.5-Omni-7B as the LMM for the verifier agent. 

\subsubsection{Generation Agent} The generation agent produces the missing modality based on guidance from the verifier agent. It selects the appropriate generator for the target modality—e.g., Stable Diffusion for images, AudioLDM 2 for audio, and large language models for text—and determines the number of candidates to balance diversity and efficiency. After generation, it collaborates with the verifier in a self-refinement loop and outputs the best-matching result. Implementation details and examples for all agents are provided in the appendix.

Across different modules, our prompt design follows a unified principle. The miner prompts are used to extract fine-grained multimodal cues from the observed modalities, such as objects, actions, scenes, and attributes; the verifier prompts assess whether candidate outputs are semantically aligned with the observed context; and the refinement prompts use verifier feedback to iteratively improve generation guidance or outputs. This unified prompt formulation allows the three modules to play complementary roles in training-free missing-modality generation. Full prompt templates are provided in the appendix.

\subsection{Self-refinement Mechanism}


From our observations, increasing the number of candidates improves generation quality but also increases inference time. Table~\ref{tab:test-time-scaling} shows that the best quality is achieved with 25 candidates, which we regard as a local optimum. However, generating 25 candidates for every instance is computationally expensive. We therefore aim to produce high-quality candidates with fewer generations.  

We formulate this as generating a sequence of subsets \(S_i\) \((i \in \{0, 1, \dots, n\})\), each containing \(m\) candidates:  
\[
S_i = \{c_1, c_2, \dots, c_m\}.
\]  
Each subset must satisfy \(Q(S_i) > Q(S_{i-1})\) for \(i > 1\), where  
\[
Q(S_i) = \max\{f(c_1), f(c_2), \dots, f(c_m)\}.
\]  
Here \(f(\cdot)\) is an LMM-based quality evaluator returning a score in \([1, 5]\), and we use Qwen2.5-Omni-7B by default.  

Given a threshold \(V_p \in [1, 5]\), if \(Q(S_i) < V_p\), the LMM provides feedback on the differences between the current candidates and the target. The generator refines its outputs accordingly to produce \(S_{i+1}\). The process terminates once \(Q(S_i) \geq V_p\), returning the best candidate. We term this process \textit{Self-refinement}.  

Self-refinement offers two benefits: (i) a smaller generation scale, since \(m\) is much less than the local-optimal (default \(m=5\)), and (ii) faster convergence, as the monotonicity constraint \(Q(S_i) \geq Q(S_{i-1})\) ensures efficient search and avoids regressions. Fig.~\ref{fig:verifier_ablation} shows that, compared with Paradigm~3, Self-refinement achieves substantially lower time costs for the same generation scale. The detailed pseudocode of the self-refinement mechanism is provided in Appendix~D.3, Algorithm~1.


\begin{table*}[t]
  \centering
  \caption{Comparison of completion accuracy and classification performance (\%) on VGGSound, COCO2014, MSRVTT, and AudioCaps under a 70\% missing rate. `/' indicates that the corresponding model is not applicable for generating the given modality.}
  {\begin{tabular}{c c|l|cc|cc|cc}
    \toprule
    Dataset & LMM & Method & \multicolumn{2}{c|}{Image} & \multicolumn{2}{c|}{Text} & \multicolumn{2}{c}{Audio} \\
     &  &  & FID $\downarrow$ & F1-score $\uparrow$ & MER $\downarrow$ & F1-score $\uparrow$ & PESQ $\uparrow$ & F1-score $\uparrow$ \\
    \midrule
    \multirow{6}{*}{VGGSound} & \multirow{3}{*}{QO} & KB~\cite{ke2025knowledge} & 130.55 & 57.96 & 0.75 & 58.22 & / & / \\
      &  & Paradigm 3 & 134.91 & 57.14 & 0.71 & 58.67 & 1.37 & 51.14 \\
      &  & AFM$^2$ (Ours) & \best 112.83 & \best 59.91 & \best 0.67 & \best 61.24 & \best 1.47 & \best 51.92 \\
    \cmidrule(lr){2-9}
      & \multirow{3}{*}{4o} & KB~\cite{ke2025knowledge} & 116.54 & 60.71 & 0.62 & 61.99 & / & / \\
      &  & Paradigm 3 & 118.91 & 60.5 & 0.69 & 61.16 & 1.49 & 52.13 \\
      &  & AFM$^2$ (Ours) & \best 97.28 & \best 62.15 & \best 0.61 & \best 63.74 & \best 1.63 & \best 52.71 \\
    \midrule
    \multirow{6}{*}{COCO2014} & \multirow{3}{*}{QO} & KB~\cite{ke2025knowledge} & 20.58 & 77.09 & 0.65 & 76.58 & / & / \\
      &  & Paradigm 3 & 20.71 & 76.03 & 0.66 & 76.06 & / & / \\
      &  & AFM$^2$ (Ours) & \best 18.06 & \best 79.04 & \best 0.59 & \best 79.88 & / & / \\
    \cmidrule(lr){2-9}
      & \multirow{3}{*}{4o} & KB~\cite{ke2025knowledge} & 16.19 & 79.37 & 0.57 & 79.61 & / & / \\
      &  & Paradigm 3 & 17.92 & 79.2 & 0.6 & 79.69 & / & / \\
      &  & AFM$^2$ (Ours) & \best 15.16 & \best 79.82 & \best 0.55 & \best 80.13 & / & / \\
    \midrule
    \multirow{6}{*}{MSRVTT} & \multirow{3}{*}{QO} & KB~\cite{ke2025knowledge} & 154.31 & 53.47 & 0.83 & 53.30 & / & / \\
      &  & Paradigm 3 & 152.27 & 52.19 & 0.88 & 53.45 & 1.31 & 52.11 \\
      &  & AFM$^2$ (Ours) & \best 146.75 & \best 56.95 & \best 0.80 & \best 55.19 & \best 1.39 & \best 54.08 \\
    \cmidrule(lr){2-9}
      & \multirow{3}{*}{4o} & KB~\cite{ke2025knowledge} & 142.4 & 60.14 & 0.79 & 59.88 & / & / \\
      &  & Paradigm 3 & 145.17 & 57.73 & 0.84 & 60.76 & 1.33 & 55.83 \\
      &  & AFM$^2$ (Ours) & \best 139.28 & \best 63.58 & \best 0.74 & \best 66.30 & \best 1.41 & \best 61.60 \\
    \midrule
    \multirow{6}{*}{AudioCaps} & \multirow{3}{*}{QO} & KB~\cite{ke2025knowledge} & 165.28 & 55.29 & 0.85 & 54.61 & / & / \\
      &  & Paradigm 3 & 171.22 & 55.36 & 0.88 & 54.08 & 1.29 & 50.29 \\
      &  & AFM$^2$ (Ours) & \best 161.39 & \best 57.18 & \best 0.82 & \best 57.47 & \best 1.40 & \best 55.83 \\
    \cmidrule(lr){2-9}
      & \multirow{3}{*}{4o} & KB~\cite{ke2025knowledge} & 158.88 & 61.73 & 0.82 & 61.65 & / & / \\
      &  & Paradigm 3 & 165.27 & 59.23 & 0.84 & 59.91 & 1.34 & 53.74 \\
      &  & AFM$^2$ (Ours) & \best 147.95 & \best 63.55 & \best 0.79 & \best 65.27 & \best 1.47 & \best 61.40 \\
    \bottomrule
  \end{tabular}}%
  \label{tab:completion_and_classification}%
\end{table*}%

\subsection{Experiments}


\subsubsection{Comparison Analysis}


We conduct comparative experiments on four datasets under a 70\% missing rate\footnote{More experiments are provided in the appendix.}. We compare AFM$^2$ against two strong baselines: Paradigm~3 and Knowledge Bridge (KB). Results are summarized in Table~\ref{tab:completion_and_classification}. Overall, AFM$^2$ achieves the best performance across datasets and modalities, indicating that the framework generalizes well.

For image completion, AFM$^2$ consistently improves both completion and downstream accuracy; for example, on VGGSound (4o), it reduces FID from 118.91 to 97.28 and raises image F1 from 60.50 to 62.15 compared with Paradigm~3. For text completion, AFM$^2$ yields lower MER and higher F1: on MSRVTT (4o), MER drops from 0.84 to 0.74, and text F1 increases from 60.76 to 66.30, while on AudioCaps (4o), MER improves from 0.84 to 0.79 and text F1 increases from 59.91 to 65.27. For audio completion, gains are smaller but consistent; on AudioCaps (4o), PESQ increases from 1.34 to 1.47 and audio F1 rises from 53.74 to 61.40, suggesting better perceptual quality and improved utility for downstream classification. We attribute these gains to our self-refinement mechanism. In KB and Paradigm~3, the generated missing modality is returned directly, without an explicit verification and re-generation step, which can make the results less robust. By contrast, AFM$^2$ verifies candidates and re-generates when necessary, bridging this gap; the consistent improvements in Table~\ref{tab:completion_and_classification} support our hypothesis.

We also note that the benefit of AFM$^2$ is not uniform across modalities. The advantage is most evident for image completion, where candidate diversity and verification are more beneficial, whereas the gains for text and audio are comparatively smaller and may come with higher inference overhead due to the additional mining and verification steps.

\begin{table}[t]
 \centering
  \caption{Components analysis on the VGGSound dataset under a 70\% missing rate.}
 \label{tab:component-ana}
 \renewcommand{\arraystretch}{1.2}
 \setlength{\tabcolsep}{3pt}
  {
  \begin{tabular}{l|cc|cc|cc}
  \toprule
    Missing Type      & \multicolumn{2}{c|}{image} & \multicolumn{2}{c|}{text} & \multicolumn{2}{c}{audio} \\
          Metrics  & FID    & F1-score  & MER   & F1-score  & PESQ & F1-score \\
    \midrule
    AFM$^2$ (4o) & 97.28 & 62.15 & 0.61  & 63.74 & 1.63  & 52.71 \\
    - \textit{w/o miner agent} & +71.12 & -0.63 & +0.09  & -2.57 & -0.31 & -2.09 \\
    - \textit{w/o verifier agent} & +13.17 & -0.38 & +0.04  & -1.03 & -0.27 & -0.26 \\
    \bottomrule
    \end{tabular}}
\end{table}

\begin{figure*}[t]
  \centering
   \includegraphics[width=1\linewidth]{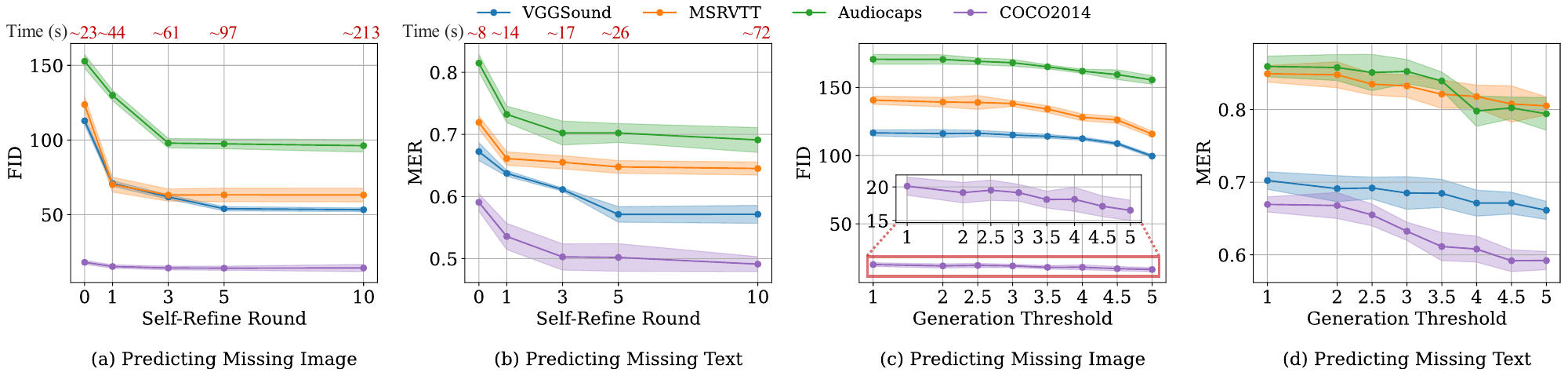}
   \vspace{-5mm}\caption{Impact of self-refinement rounds (0, 1, 3, 5, 10) and generation threshold values (1.0–5.0) on the quality of missing modality generation under a 70\% missing rate.}\vspace{-2mm}
   \label{fig:verifier_ablation}
\end{figure*}

\subsubsection{Ablation Study}

We further analyze how individual components, the number of self-refinement rounds, and the generation threshold affect both generation quality and downstream classification in AFM$^2$. Table~\ref{tab:component-ana} reports an ablation on VGGSound at a 70\% missing rate. The miner agent is the primary driver of performance: in image generation, removing it markedly increases FID (\textcolor{red}{+71.12}). By comparison, removing the verifier agent produces a moderate decline (FID \textcolor{red}{+13.17}). These results indicate that both agents matter, with the miner agent exerting the larger effect.  

We next study two verifier factors: self-refinement rounds and the generation threshold (Fig.~\ref{fig:verifier_ablation}). Under a 70\% missing rate, consistent patterns appear across all four datasets for FID (image) and MER (text). In Fig.~\ref{fig:verifier_ablation}(a)--(b), increasing the number of rounds generally improves performance, with optimal results typically between 1 and 5 iterations; gains beyond this range are marginal. Jointly with Table~\ref{tab:test-time-scaling}, rounds of 1, 3, and 5 provide a more favorable time--quality trade-off than Paradigm~3 (QO) under the same total candidate budget, with the clearest advantage observed in image completion and more modest gains for text and audio. To further examine the scaling behavior of AFM$^2$, we additionally vary the number of generated candidates while keeping the self-refinement procedure unchanged. As shown in Appendix Table~VI, AFM$^2$ generally benefits from larger candidate pools across image, text, and audio completion. For example, under the 4o setting, increasing the candidate number from 5 to 25 improves image FID from 97.28 to 93.27, text MER from 0.61 to 0.59, and audio PESQ from 1.63 to 1.83, with corresponding downstream F1 improvements across all three modalities. Meanwhile, the gains gradually saturate as the candidate number increases, while the inference cost grows substantially, indicating a clear quality--efficiency trade-off. For thresholds (Fig.~\ref{fig:verifier_ablation}(c)--(d)), values below 3.5 degrade performance, likely due to lenient acceptance of low-quality candidates, whereas moderate thresholds (e.g., 4.0--4.5) provide a more balanced compromise between quality control and stability.

\subsection{Visualization}

\begin{figure*}[t]
  \centering
   \includegraphics[width=0.9\linewidth]{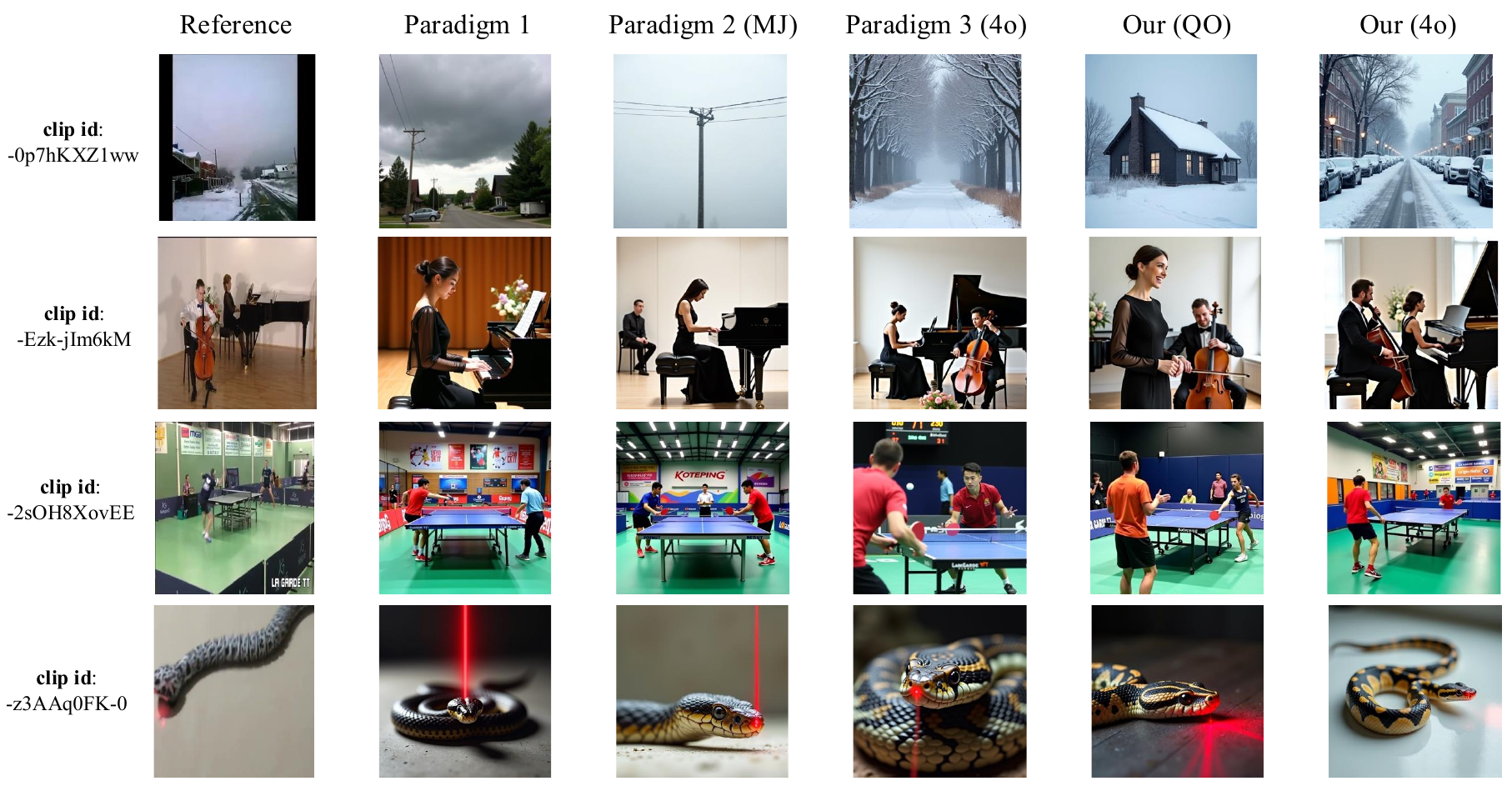}
   \vspace{-2mm}\caption{Visualization of missing image generation results from different paradigms on the VGGSound dataset.}\vspace{-2mm}
   \label{fig:vggsound-1}
\end{figure*}

\begin{figure*}[t]
  \centering
   \includegraphics[width=0.9\linewidth]{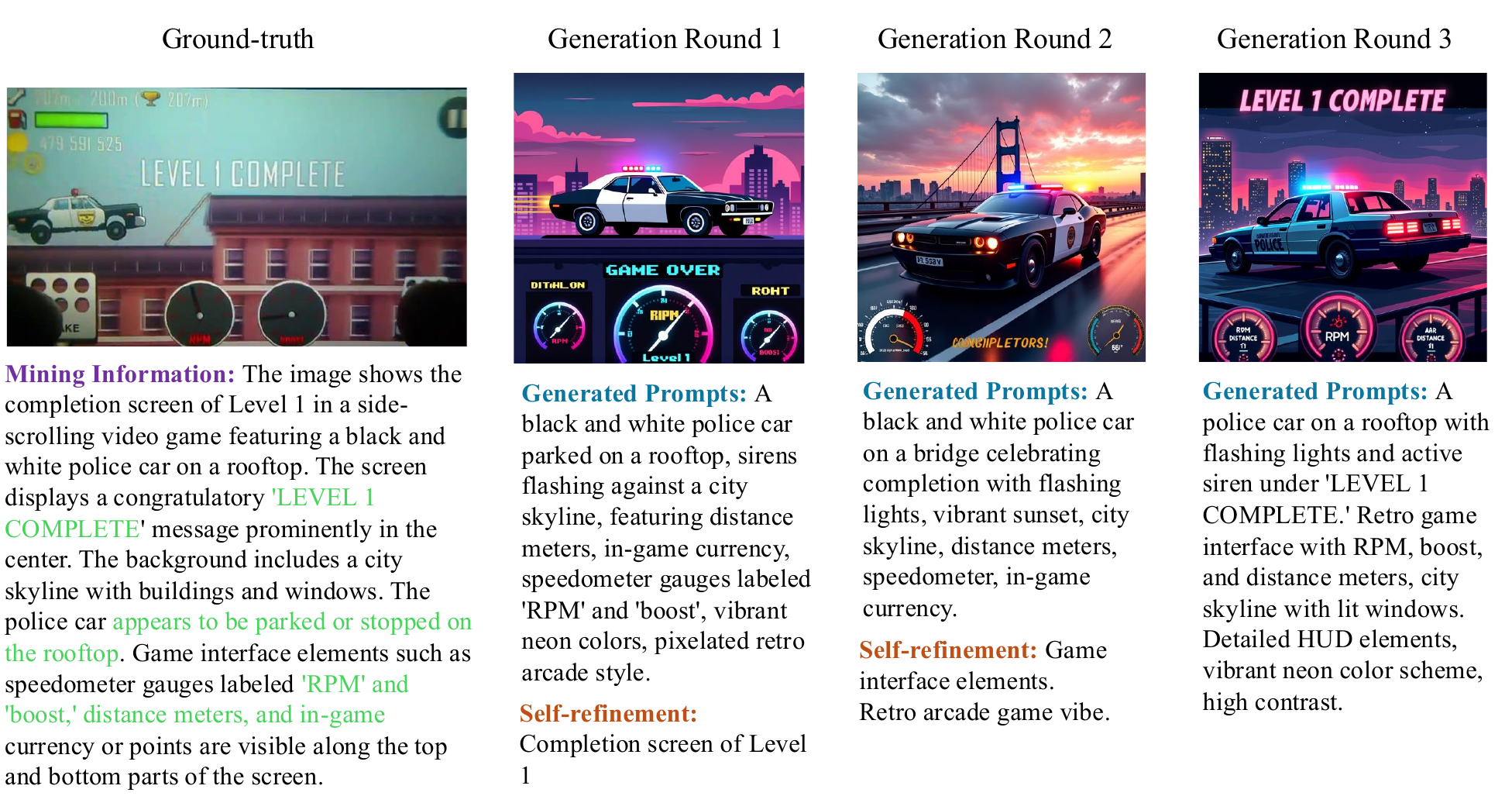}
   \vspace{-2mm}\caption{Visualization of the self-refinement mechanism results.}\vspace{-2mm}
   \label{fig:vggsound-regen-2}
\end{figure*}

Figures~\ref{fig:vggsound-1} and \ref{fig:vggsound-regen-2} present the qualitative results for missing image restoration and the visualization of the self-refinement process, respectively. The results show that methods incorporating the miner and filter modules (Paradigm~3 and Ours) produce higher-quality reconstructions than Paradigm~1 and Paradigm~2. Moreover, approaches leveraging stronger models, such as GPT-4o, can capture finer details. These observations are consistent with our claims. In addition, Fig.~\ref{fig:vggsound-regen-2} demonstrates that our method can effectively review the deficiencies of the current outputs and generate informative feedback for iterative refinement. Overall, the self-refinement mechanism substantially improves the quality of missing modality generation in a training-free setting.

\section{Discussion}

\noindent\textbf{Q1: How far are we from generating missing modalities using foundation models in a training-free manner?}  
Multimodal foundation models hold promise for plug-and-play missing modality generation, with applications spanning healthcare, scientific discovery, and daily-life assistance. This work offers a systematic evaluation of current progress toward this goal. Our results highlight two key challenges. First, generating missing audio remains substantially more difficult than vision or text. Second, existing models struggle to capture deep cross-modal relationships—particularly in vision-text settings. While incorporating an explicit miner module partially alleviates these issues, most models still lack output verification mechanisms, contributing to hallucinations and inconsistencies. Addressing these limitations will require enhanced semantic reasoning across modalities and robust validation strategies. We hope our findings encourage future efforts toward more reliable and unified large multi-modal models.

\noindent\textbf{Q2: What enables a pre-trained generator to produce the desired missing modality?}  
Our analysis shows that the introduction of miner and ranking modules substantially improves generation quality. However, these improvements are achieved indirectly—by refining generation instructions—rather than by enabling the generator to truly understand what should be emphasized or generated. This limitation stems from the prevalent reliance on text-only prompts. Prior studies~\cite{wang2024emu3, chen2025janus, hurst2024gpt} suggest that integrating additional modalities can enhance generation performance. We argue that empowering generators to jointly interpret both instruction prompts and external multimodal cues is essential for advancing future generation capabilities.

\noindent\textbf{Limitations.}  
While our work systematically investigates the training-free capabilities of foundation models in generating missing image, text, and audio modalities, several limitations remain. First, we focus exclusively on inference-time paradigms and do not explore training-efficient adaptation methods such as LoRA~\cite{hu2022lora} or prompt tuning~\cite{lester2021power}, which may offer further gains under constrained computational budgets. Second, our evaluation is limited to three modalities, leaving open questions about generalizability to others such as video, tabular data, or sensor streams—modalities of growing importance in real-world applications. Third, although our miner and verifier modules improve generation via indirect instruction refinement, current generators still lack the capacity for joint reasoning over prompts and external semantic inputs. Future work will extend our framework to broader modalities and incorporate lightweight adaptation to enhance versatility and generalization.

\section{Conclusion}

We systematically investigate the capability of foundation models in missing modality generation, considering both generation accuracy and downstream task adaptability. To this end, we construct three representative paradigms comprising 42 variants. Extensive experimental results demonstrate that mining information from available modalities and ranking generated candidates play a critical role in improving generation quality. Motivated by these findings, we propose an agentic framework featuring automated information mining and a self-refinement mechanism. The framework dynamically derives mining strategies based on the observed modalities and evaluates the quality of generated candidates for iterative refinement. Comparative and ablation studies show that our method leads to substantial improvements in the quality of generated missing modalities.




\bibliographystyle{IEEEtran}
\bibliography{main}

\end{document}